# Performance Analysis of Best suited Adaptive Equalization Algorithm for Optical Communication

Tarek Hasan-Al-Mahmud, M.Mahbubur Rahman, Sumon Kumar Debnath

**Abstract**— Fiber optics is one of the highest bandwidth communication channel types in the current communication industry. The paper is to analyze a typical optical channel and perform channel equalization using an adaptive modified DFE with Activity Detection Guidance and Tap Decoupling algorithm. Evaluation can be made on the employment of the DFE algorithm and with enhancements, like Fractionally-Spaced equalization and Activity Detection Guidance, to improve its stability, steady-state error performance and convergence rate. The successful implementation of the Adaptive FS-DFE with ADG and TD technique offers an excellent alternative to linear equalization, which is known to be of little benefit for optical channels because of exorbitant noise enhancement. The FSE technique, when combined with the DFE, would offer improved effectiveness to amplitude distortion. As the impulse response of a typical optical link would have regions that are essentially zero, the employment of the activity detection scheme with Tap Decoupling would further enhance the steady-state error performance and convergence rate.

**Index Terms**— Adaptive Fractionally spaced, Modified Decision Feedback Equalizer, Activity Detection Guidance, Tap-Decoupling,

——————————— ◆ ———————————

## 1 INTRODUCTION

It is well known that communication speed is increasing rapidly over the years. Long haul transmission in optical communications usually consists of several keys area to ensure the data transfer is error-free. These key areas include the modulation method, type of amplifier used, and the error correction scheme and dispersion compensation. In optical communications, high-speed data transfer is often limited by signal distortion, which is mainly caused by the broadening of pulses that result in Intersymbol Interference (ISI).

In optical communications, the main source of distortion comes from pulse dispersion. Pulse dispersion can be separated into two main groups, namely: Intermodal Dispersion and Intramodal Dispersion. Intramodal or chromatic dispersion occurs in all types of fibers. Intermodal or modal dispersion occurs only in multimode fibers. Each type of dispersion mechanism leads to pulse spreading. As a pulse spreads, energy is overlapped.

————————————————

● *Tarek Hasan-Al-Mahmud is with the Computer Science and Telecommunication Engineering Department, Noakhali Science and Technology University, Noakhali, Bangladesh.*
● *M.Mahbubur Rahman is with the Information and Communication Engineering Department, Islamic University, Kushtia. Bangladesh.*
● *Sumon Kumar Debnath is with the Computer Science and Telecommunication Engineering Department, Noakhali Science and Technology University, Noakhali, Bangladesh.*

The dispersion also limits the transmission distance. Regenerative repeaters, which reconstruct and boost the signal shape and power are needed more frequently if the dispersion is severe. Such repeaters are expensive to build and install, especially when the optical link runs under the seabed in cross-ocean communications links. The aim of this paper is first to evaluate to the causes of dispersion in a typical optical link and its limiting effects to the possible communication capacity. The Least Mean Square (LMS) algorithm was chosen for its robustness and computational simplicity. Therefore, the Modified Decision Feedback Equalizer (DFE) combined with Fractionally Spaced Equalizer and Activity Detection Guided with Tap Decoupling was chosen as the most suitable equalizer structure for an optical channel that suffers from amplitude distortion. An accurate way of analyzing the behavior of an adaptive DFE algorithm and the impulse response of an optical channel is evaluated by using the simulation package Matlab_07. It allows replicating specific and realistic impulse responses of a particular channel and the performance of a certain adaptive algorithm in indirect mode approach. Evaluation of the algorithm can be achieved in terms of channel estimation, asymptotic performance, convergence rate and squared difference of error performance.

## 2 TYPES OF EQUALIZERS SUITABLE FOR AN OPTICAL CHANNEL

In a practical optical communication channel, amplitude distortion is one of the major detrimental effects. Therefore, the investigation and research into the Decision Feedback Equalizer, which is a non-linear equalizer and capable of superior performance in amplitude distorted channels, would be very beneficial and relevant to the application in optical communications. When implementing the DFE structure, enhancements like the Fractionally Spaced Equalization, which makes the equalizer more robust to amplitude distortions can also be considered.

### 2.1 Fractionally-Spaced Equalizer (FSE)

Fractionally Spaced Equalizer (FSE), which is based on sampling the incoming signal at least as fast as the Nyquist rate. Fractionally-spaced equalizers have taps that are spaced closer than conventional adaptive equalizers, and with a sufficient number of taps, it is almost independent of the channel delay distortion. It means that the equalizer can negate the channel distortion without enhancing the noise [1]. Given the above properties, the FSE technique is a highly desirable application since it minimizes noise enhancement. With appropriately chosen tap spacing; the FSE can be configured as the excellent feedforward filter.

When we combine the DFE with FSE technique, we can expect an equalizer with the following qualities:

1. Minimize noise enhancement
2. Excellent amplitude distortion performance.

### 2.2 Decision Feedback Equalizer

In a Decision Feedback Equalizer, both the feedforward and feedback filters are essentially linear filters. It is a non-linear structure because of the non-linear operation in the feedback loop (decision threshold); its current output is based on the output of previous symbols. The reason for choosing DFE over linear equalizer is that the latter's performance in channel that exhibit nulls is not effective. In contrast, the decision feedback equalizer has zero noise enhancements in the feedback loop. The above DFE structure has $N_1 + N_2 + 1$ feedforward taps and $N_3$ feedback taps. [2] The output of the equalizer is given by:

$$\hat{d}_k = \sum_{n=-N_i}^{N_2} c_n^* y_{k-n} + \sum_{i=1}^{N_2} F_i d_{k-i} \qquad (1)$$

where $c_n^*$ is the tap gain and $y_n$ is the input for the forward filter, $F_i^*$ is the tap gain for the feed back filter and $d_i(i<k)$ is the previous decision made on the detected signal. When implementing a filter such as the DFE, where the adjustment term is based partly on past decision of symbols, there exists a probability that the decision made by the decision threshold device may be wrong. The occurrence of one error in the DFE will cause a burst of new errors. Its error probability can be kept under 'useable' condition with some conventional considerations: 1. Making suitable system choices, 2. Careful consideration of algorithm implementation, 3. Keeping length of feedback filter short enough, 4. Use of Tomlinson-Harashima precoding and 5. Detection of error correction scheme. A point to note, though, is that the error probability of the DFE cannot be improved by error correcting coding (for example, channel coding). This is because an immediate decision is required at the output of the decision threshold device. The Tomlinson-Harashima precoding and Error correction scheme would be beneficial to remove this problem. With Tomlinson-Harashima (TH) precoding [3], the feedback filter (FBF) of the decision feedback equalizer is implemented at the transmitter to avoid the effects of error propagation in which a detection error advances through the FBF. The principle behind MDFE this is that since the coefficient of the FBF can be initialized using an estimate of the channel impulse response, it is only required to initialize and train the FFF for the start-up. As a result, the MDFE can reach steady-state faster than conventional DFE [4].

### 2.3 Adaptive Equalization Approach

Fundamentally, there are two basic approaches in adaptive equalization of an unknown channel, the direct and indirect approach. The indirect approach is illustrated with Figure 1 below:

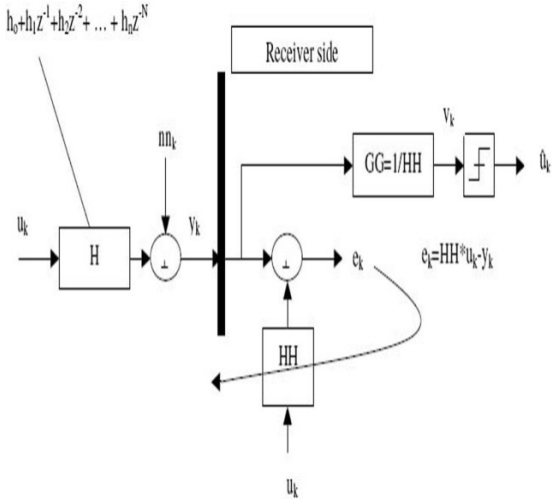

Figure 1   Indirect Equalization Approach

The indirect approach, also known as equalization via channel estimate, uses the training sequence to adapt the filter coefficients to replicate the response of the unknown channel. The filter coefficients would then converged close to the optimum solution before the end of the training sequence. After which, GG is updated and the actual data received would then be pass through GG and subsequently a decision threshold device to output the actual data sent. For this paper the indirect approach is adopted as the approach to adaptive equalization.

## 3 IMPULSE RESPONSE OF A TYPICAL OPTICAL CHANNEL

The impulse response of a typical optical channel is known to have a cosine-squared pulse shape. The impulse response of a typical optical channel is given by [5]:

$$h(t) = \begin{cases} \frac{2}{\tau}\cos^2\frac{\pi t}{\tau}, & -\frac{\tau}{2} \leq t \leq \frac{\pi}{2} \\ 0, & elsewhere \end{cases} \quad (2)$$

And the baseband frequency response is given by [5]:

$$h(f) = \frac{\sin \pi f t}{\pi f t}\left[\frac{1}{1 - f^2\tau^2}\right] \quad (3)$$

From Equation 3, it can be observed that the channel will have nulls wherever $\sin \pi f t = 0$ (except at $f = 0$ and $f = 1/\pi f t$). This also gives the motivation for the use of the Activity Detection Guidance technique.

## 4 ACTIVITY DETECTION GUIDANCE

Activity detection guidance technique is a method of detecting active taps in a communication channel. By implementing a technique capable of detecting active taps in the channel, non-active taps can be excluded in the estimation of the channel response. This relieves the computational burden of the LMS algorithm, as well as to give a better convergence rate and asymptotic performance. The detection of the 'active' taps of a time-invariant channel is governed by the equation [6]:

$$C_{k,n} = \frac{\left(\frac{1}{N}\sum_{i=1}^{N}iu.y_{i-k+1}\right)^2}{\frac{1}{N}\sum_{i=1}^{N}(y_{i-k+1})^2} \quad (4)$$

where i=time index, k=tap index, and N is the number of input samples. $C_k$ is known as the activity measure. In order to determine a tap to be active, the value of the activity measure, $C_k$, must be above a certain threshold. This activity threshold is given by:

$$C_{k,N} > \sum_{i=1}^{N}\frac{(iu)^2 \cdot \log(i)}{i} \quad (5)$$

The accuracy improves with increasing N. When a tap is detected as inactive, it will be excluded in the estimation of the active taps.

### 4.1 Activity detection guidance with Tap Decoupling

Modifications to the activity measure have to be made to reduce the tap coupling effect [6].

$$CC_k = \sum_{i=1}^{N}\frac{[(iu_i - hiu_i + H_kY_{i-k+1}).(Y_{i-k+1})]^2}{\sum_{i=1}^{N}(Y_{i-k+1})^2} \quad (6)$$

This modification would reduce the contribution of any adjacent active tap to the perceived activity of the actual tap being considered.

## 5 SIMULATION RESULTS AND DISCUSSIONS

This section will present the results of the simulations performed using Matlab_07, based on the concepts described in earlier sections. The results were analyzed and presented using the parameters set to evaluate the performance of an adaptive equalizer, namely: 1. Channel adaptation, 2. Asymptotic performance, 3. Convergence rate and 4. Squared difference of input to Channel and output of Equalizer. The channel has a total of 2 active taps and a total tap length of 7. The noise interference applied was a zero mean white Gaussian signal, with a variance of 0.1. The adaptation step size was set at 0.005. The simulations involved the following:

- Fractionally-spaced Modified Decision Feedback Equalizer using standard LMS with Activity Detection Guidance
- Fractionally-spaced Modified Decision Feedback Equalizer using standard LMS with Activity Detection Guidance and Tap Decoupling

## 5.1 Fractionally spaced Decision Feedback Equalizer with Activity Detection Guidance

Activity Detection Guidance scheme was incorporated into the FS-DFE. Its purpose is to detect the non-active taps, disregard them in the calculation of the tap weight adaptation and set them to zero. By disregarding the inactive taps, convergence rate is most certainly to improve the asymptotic performance. To compare the simulated channel impulse response and FS-DFE with ADG estimated channel impulse response both channel estimation graphs are presented respectively

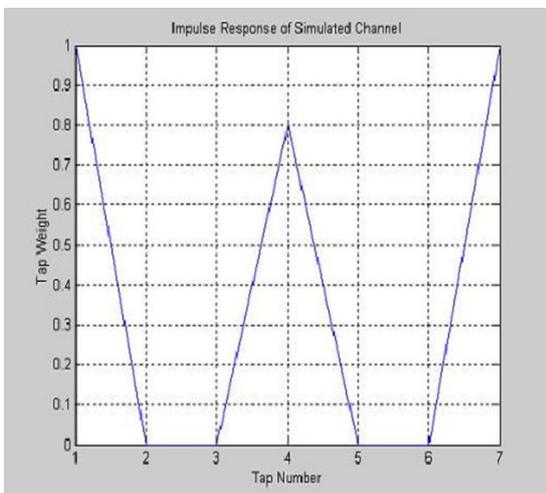

Fig. 2 Simulated channel impulse responses

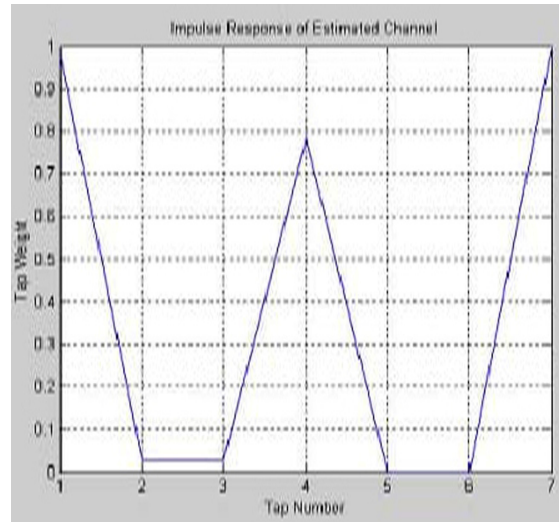

Figure 3  Estimated Channel Impulse Response of AFS-MDFE-ADG

From figure3 it can be observed that the FS-DFE with ADG was capable of detecting the magnitudes and positions of the active taps in the channel and was able to ignore the estimation of the inactive taps partially as expected.

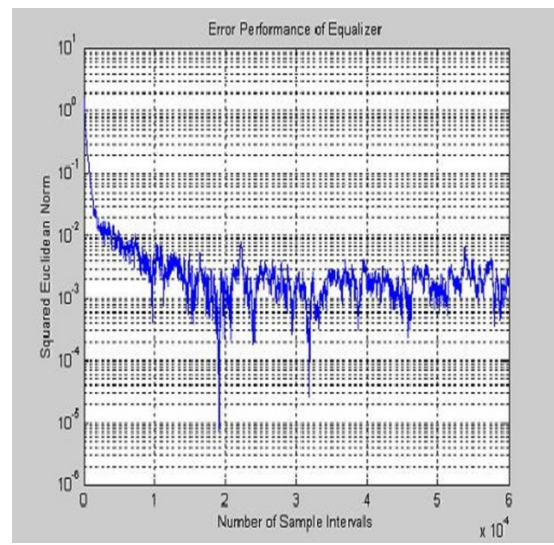

Figure 4  Asymptotic performance of AFS-MDFE-ADG

From figure4 it can be observed that the convergence rate was faster and the asymptotic performance is $10^{-3}$. This is a significant improved value even though the non-active taps were not ignored in the tap weight estimation.

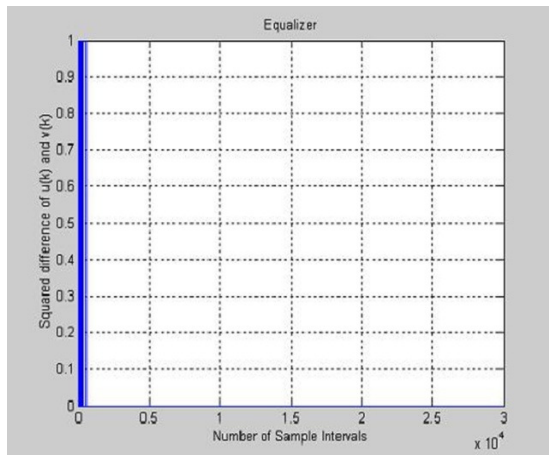

Figure 5   Output of equalizer of AFS-MDFE-ADG

From figure5 it can be observed that the output of the equalizer has the burst of errors in the beginning is still existing but considerably reduced and the number of errors encountered was notably lesser.

### 5.2 Fractionally spaced Decision Feedback Equalizer with Activity Detection Guidance and Tap Decoupling

It was noted in the previous sections that the FS-DFE with ADG showed considerable improvement in convergence rate and asymptotic performance over the FS-DFE alone. However, the true number of active taps is not detected accurately and the inactive taps still suffer from the noisy estimates that are common in any communication channel. Using the principle behind the Activity Detection Guidance scheme, the tap coupling effect can be overcome with some modifications to the activity measure. The fine results can be seen below.

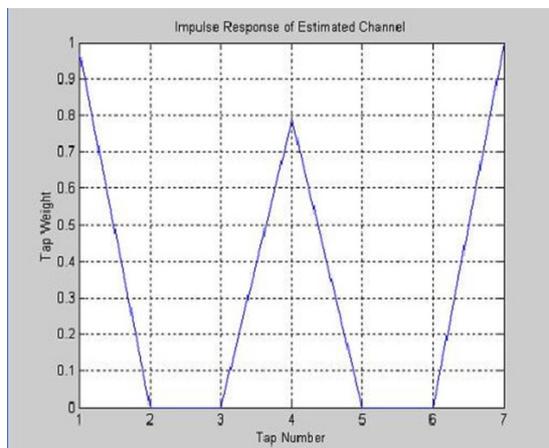

Figure 6   Estimated Channel Impulse Response of AFS-MDFE-ADG with TD

Figure6 shows the impulse response of the estimated channel is almost an exact replica of the simulated channel. The magnitude and positions of the active taps were detected and adapted closely and the inactive taps were bound to zero.

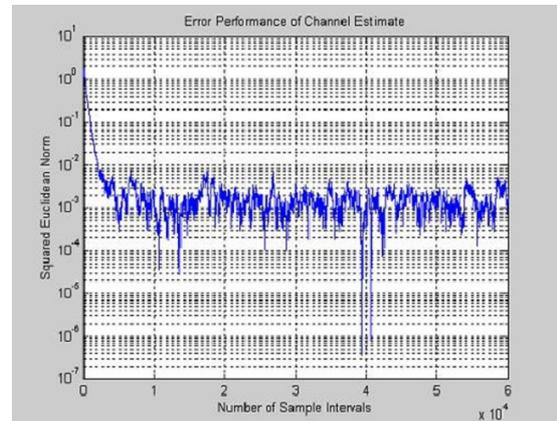

Figure 7   Asymptotic performance of of AFS-MDFE-ADG with TD

Figure7 shows the asymptotic performance of the equalizer to be $10^{-3}$. This is the almost the same as the FS-DFE with ADG, however, the convergence rate of the FS-DFE with ADG and TD improved by 2 fold as compared to the FS-DFE with ADG.

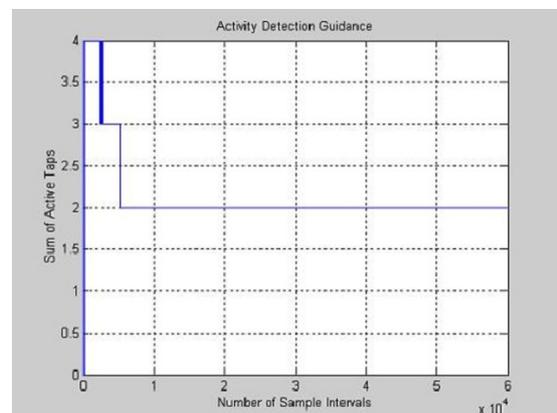

Figure 8   Active Tap Count of AFS-MDFE-ADG with TD

Figure8 shows the true number of active taps being detected. The slight 'jitters' at the initial stage is most certainly due to noise as the adaptive filter converges to measure the correct number of active taps. This proves that the FS-DFE with ADG and TD has successfully been implemented and working effectively.

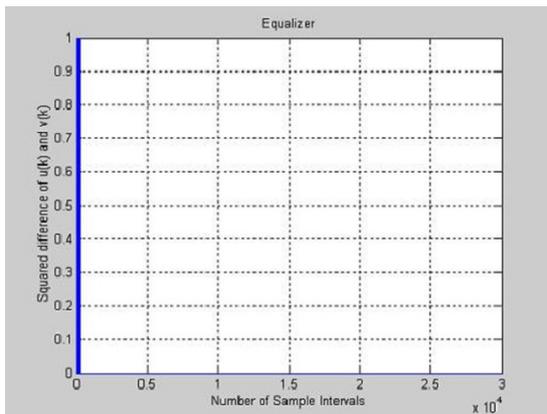

Figure 9   Equalizer Output

Figure9 shows the accuracy of the output of the equalizer. Besides the burst of errors in the initial stage, there were zero errors for the rest of the data. This shows that the equalizer has adapted effectively and is mitigating the distortion effects of the channel successfully.

### 5.3   Comparison between FS-DFE, FS-DFE with ADG, and FSDFE with ADG and TD

The FS-DFE with the activity detection and tap decoupling enhancements has the best performance throughout the parameters set to assess an adaptive equalizer. The convergence rate and asymptotic performance can be appreciated more clearly from figure 10 below.

Blue – FS_DFE, Green – FS_DFE with ADG, Red – FS_DFE with ADG and TD.

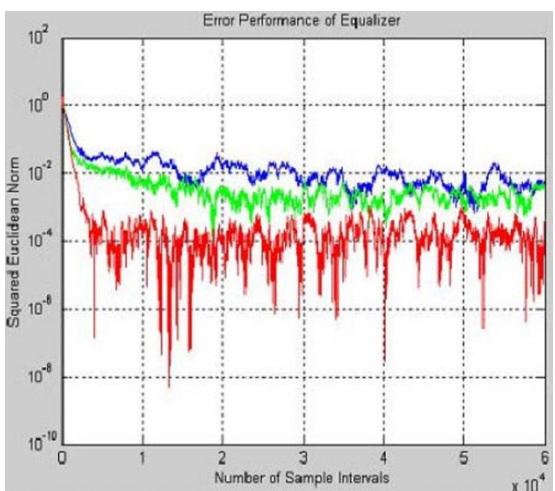

Figure 10   Comparison of Asymptotic performance

## 6 CONCLUSION AND RECOMMENDATION FOR FUTURE WORKS

The objective of this paper is to develop a suitable adaptive equalization technique to mitigate the effects of ISI and dispersion in a typical optical communication channel. With the successful development of the adaptive modified Decision Feedback Equalizer with activity detection guidance and tap decoupling, it offers an excellent alternative to the existing equalization techniques available in the optical communication. The adaptive equalizer was implemented using the Least Mean Square (LMS) technique, using stochastic gradient adaptation, for the indirect equalization of the unknown channel. This led to a faster convergence rate, as only the active taps need to be adapted, and better asymptotic performance as shown clearly. In the nutshell, this work has demonstrated the successful implementation of an adaptive modified Decision Feedback Equalizer with ADG and TD in a typical optical communication channel. The theoretical research and findings were successfully implemented and proven. This design approach is a promising alternative for equalization in optical communications.

A suggestion for future expansion in this area of research is to develop an adaptive DFE using the recursive least squares algorithm on a lattice structure to provide faster convergence.

### REFERENCES

[1]  R. D. Gitlin and S. B. Weinstein, "Fractionally-spaced equalization: An improved digital transversal equalizer", The Bell System Technical Journal, Vol. 60, No. 2, February 1981.
[2]  Theodore S. Rappaport, "Wireless Communications – Principles and Practice" 2nd edition, Prentice Hall Communications Engineering and Emerging Technologies Series, Upper Saddle River, New Jersey, 2002.
[3]  John E. Smee and Stuart C. Schwartz, "Adaptive Compensation Techniques for Communications Systems With Tomlinson–Harashima Precoding", IEEE transactions on communications, Vol. 51, No. 6, June 2003.
[4]  Heechoon Lee, Sukkyun Hong, Yong-Hwan Lee and Kwang-Bok Lee, "Fast training of fractionally-spaced modified decision feedback equalizer in slow frequency selective fading channels", IEEE 55th , Volume: 4 , 6-9 May 2002
[5]  B.L.Kaspers, "Equalization of Multimode Optical Fiber Systems", Bell System Technical Journal, September 1982.
[6]  John Homer, Iven Mareels, Robert R. Bitmead, Bo Wahlberg and Fredrik Gustafsson, "LMS Estimation via

Structural Detection", IEEE Transactions on signal processing, Vol. 46, No. 10, October 1998.